\documentclass[prd,floats,nopacs]{revtex4}
\usepackage{amsmath,amsfonts,amssymb,amsthm}
\usepackage{graphicx, epsfig}

\newcommand{\sDelta}{{\scriptstyle\Delta}}

\newcommand{\ii}{\mathrm{i}}

\newcommand{\diff}{\mathrm{d}}

\newcommand{\bigO}{\mathit{O}}
\DeclareMathOperator{\sgn}{sgn}

\begin{document}
\title{Peculiar seasoning in the neutrino day-night asymmetry: where and when to look for spices? }
\author{Oleg~G.~Kharlanov} \email{okharl@mail.ru}
\author{Andrey~E.~Lobanov} \email{lobanov@phys.msu.ru}
\affiliation{Department of Theoretical Physics, Faculty of Physics, Moscow State University, 119991
Moscow, Russia}

\begin{abstract}
We analyze the peculiar seasonal effects in the day-night asymmetry of
solar neutrinos, namely those connected with the neutrino nighttime flux anomaly
near the winter solstice. We show that, for certain placements of the neutrino
detector, such effects may be within the reach of next-generation detectors.
\end{abstract}

\maketitle

In our recent paper \cite{1}, we have shown that the Earth matter effects for both
Beryllium-7 ($E = 0.862$ MeV) and Boron-8 ($E \sim 4 - 15$ MeV) solar neutrinos
can be efficiently described using analytical techniques. Indeed, the parameters
of the Earth's internal structure and the typical oscillation lengths of solar
neutrinos are such that not only the oscillation probabilities for neutrinos at
a given nadir angle $\Theta_{N}$ can be calculated analytically using the valley-cliff
approximation \cite{2},
\begin{equation}\label{valleyCliff}
      P_{\nu_e}(\Theta_{\text{N}}; E) \approx \begin{cases}
      \frac12 +  \frac12 \cos2\theta_{\text{Sun}}\cos2\theta_0, & \Theta_{\text{N}} > \pi/2 \text{ (day)}, \\
      \frac12 +  \frac12 \cos2\theta_{\text{Sun}}
      \bigl\{\cos2\theta_n^- + 2\sin2\theta_0\sum\limits_{j=1}^{n-1}
      \sDelta\theta_j\cos2\sDelta\psi_{n,j}
      \bigr\},
      &
      \Theta_{\text{N}} < \pi/2 \text{ (night)},
      \end{cases}
\end{equation}
\noindent but one can also find an analytical expression for the electron neutrino observation
probability averaged over a year-long observation term
\begin{equation}\label{yearAvg_def}
      \langle P_{\nu_e}(\,^{\text{night}}_{\text{day}};E) \rangle_{\text{year}} \equiv \int\limits_{\text{1~year}}\frac{w(t)\diff{t}}{0.5\text{~year}}\;
      \vartheta(\pm \pi/2 \mp \Theta_{\text{N}}(t)) \; P_{\nu_e}(\Theta_{\text{N}}(t); E).
   \end{equation}
where $w(t)$ is a smooth weighting function. In Eq. \eqref{valleyCliff}, $\theta_{0}$ is the vacuum mixing
angle (we use the two-flavor approximation), while $\theta_{\mathrm{Sun}}$ is the effective mixing
angle in the solar core \cite{3}. The nighttime solar neutrino with energy $E$ is
assumed to cross $n = n(\Theta_{N})$ interfaces on its way through the Earth to the detector
and $\theta_j^\pm$ are the effective mixing angles in the Earth's medium after/before
the $j$th interface; $\sDelta\theta_j \equiv \theta_j^+ - \theta_j^-$ are the effective mixing angle jumps. Moreover,
$\theta_n^-$ is the effective mixing angle in the crust under the neutrino detector,
while $\sDelta\psi_{n,j}$ is the oscillation phase incursion between the $j$th crossing point
and the detector, $\pi$ per every oscillation length in the medium \cite{1}.

The opportunity to analytically average the electron neutrino observation
probability \eqref{valleyCliff} over the solar motion $\Theta_{N}(t)$ during the whole year stems from
the following. As the Sun ascends and descends, the oscillation phases $\sDelta\psi_{n,j}(\Theta_{N}(t),E)$ change by much more than $\pi$, thus the cosines $\cos2\sDelta\psi_{n,j}$  in Eq. \eqref{valleyCliff} 
are rapidly oscillating functions of time. Then, the time average of every of
such oscillating contributions to the nighttime neutrino observation probability \eqref{yearAvg_def} can be evaluated using the stationary phase (saddle point) approximation
\begin{equation}\label{statPhase}
       \int\limits_a^b f(t) e^{\ii \lambda S(t)} \diff{t} = \sum\limits_{j=1}^{p}
       \sqrt{\frac{2\pi\ii}{\lambda S''(t_s)}} f(t_s) e^{\ii\lambda S(t_s)}
       + \left.\frac{f(\tau)e^{\ii\lambda S(\tau)}}{\ii\lambda S'(\tau)}\right|_a^b + \bigO(\lambda^{-3/2}),
       \end{equation}
\noindent where functions $f(t)$ and $S(t)$ are smooth on a segment $[a, b]$ containing $p > 0$
isolated non-degenerate stationary points $t_{s} \in(a, b)$ such that $S'(t_{s}) = 0$,
$S''(t_{s}) \neq 0$, and $\lambda \rightarrow +\infty$ \cite{4}. In the case of nighttime solar neutrinos, the
oscillation phase for the $j$th interface $\lambda S(t)\equiv 2\sDelta\psi_{n,j}$ obviously achieves stationary
points at midnights, while the boundary terms in Eq. \eqref{statPhase} vanish \cite{1}.
Thus, the time integral \eqref{yearAvg_def} reduces to a sum over 365 midnights. This sum can
also be replaced by an integral, and the latter once again demonstrates isolation
of the stationary points. These two points are the summer and the winter
solstices, corresponding to the lowest and the highest midnight solar positions.
Finally, for a smooth normalized weighting function $w(t)$ (see definition \eqref{yearAvg_def}),
the year-average day-night asymmetry is \cite{1}
\begin{multline}\label{deltaP_year}
       \langle P_{\nu_e}(\text{night})-P_{\nu_e}(\text{day}) \rangle_{\text{year}} \approx \frac12 \cos2\theta_{\text{Sun}} (\cos2\theta_n^-
       -\cos2\theta_0)
                   + \cos2\theta_{\text{Sun}}\sin2\theta_0
                   \\ \times\sum\limits_{j=1}^{n-1} \sDelta\theta_j
                   \sum\limits_{s=\pm1}w(t)
                   \frac{\vartheta\big(r_j - r_n\sin(\chi+s\varepsilon)\big)}
                   {2\pi\sqrt{\sin\varepsilon \cos\chi \sin(\chi+s\varepsilon)}}
                   \frac{\sqrt{r_j^2/r_n^2 - \sin^2(\chi+s\varepsilon)}}{\sDelta m^{2}
                   L_{n,j}^{\text{solstice}}/4E}
                   \cos\left\{2\sDelta\psi_{n,j}^{\text{solstice}} + s'(s-1) \frac{\pi}{4}\right\}.
   \end{multline}
\noindent In the above expression, $s = \pm 1$ indexes the two solstices, $\Theta_{N}= \lambda+s\varepsilon$ is
the solar nadir angle at the solstice midnight, $\chi$ is the latitude of the detector,
$\varepsilon = 23.5^{\circ}$ is the Earth's axial tilt. The radii $r_{j}$ of the interfaces between the
spherical layers inside the Earth, for which the Heaviside $\vartheta$ function is nonzero,
correspond to those interfaces which are actually crossed by the neutrino ray
at the solstice midnight, at distance $L^{\mathrm{solstice}}_
{n,j}$ from the detector. Each interface
enters the sum in the Eq. \eqref{deltaP_year} twice, once referring to the neutrino going into
the interface and once out of it; $s'\equiv\sgn\{L^{\mathrm{solstice}}_
{n,j}-r_{n}\cos(\chi+s\varepsilon)\}$   is positive
for the `entry' points and negative for the `exit' points. The oscillation phase
incursion $2\sDelta\psi_{n,j}^{\mathrm{solstice}}$
is taken between the point where the neutrino crosses the
$j$th interface and the detector (the $n$th interface).

One easily observes that the contribution to the seasonal average of the day-night
asymmetry contains the terms localized around the solstices, with quite
peculiar properties:

a) they are sensitive to the seasonal distribution of the observations, i.e.,
the weight $w(t)$; for an experiment carried out around the solstices, they
undergo an amplification inversely proportional to the observation term;

b) the winter solstice $(s =-1)$ contribution formally becomes infinite at the
Tropic $\chi=\varepsilon$, while the summer contribution remains finite;

c) the magnitude of the $j$th term does not depend on how often during the
year the Sun descends low enough to shine through the $j$th interface; this
is a direct manifestation of the time localization of the contributions;

d) they oscillate with the neutrino energy $E$, as well as with the radii $r_{j}$ of
the interfaces; deeper interfaces produce faster oscillations with $E$.

Thus, observation of the peaks on the day-night asymmetry energy spectrum
could, in principle, provide a way to measure the neutrino mass-squared difference
$\sDelta m^{2}$ and/or the radii $r_{j}$ . Moreover, if one, say, takes into account the
day-night effect data for December and January, the number of neutrino events
will be $6$ times less than that for a year-long observation and this will spoil (in
terms of the statistical error) the observation of the constant contribution to
day-night effect by the factor of
$\sqrt{6}$.
On the other hand, every oscillating contribution
to Eq. \eqref{deltaP_year} will become $\approx 6$ times larger over such a small observation
period, hence, observing such contributions `around the Christmas' is $\sqrt{6}$ times
more efficient than doing it homogeneously throughout the year.

\begin{figure}[h]
\includegraphics[width=14cm]{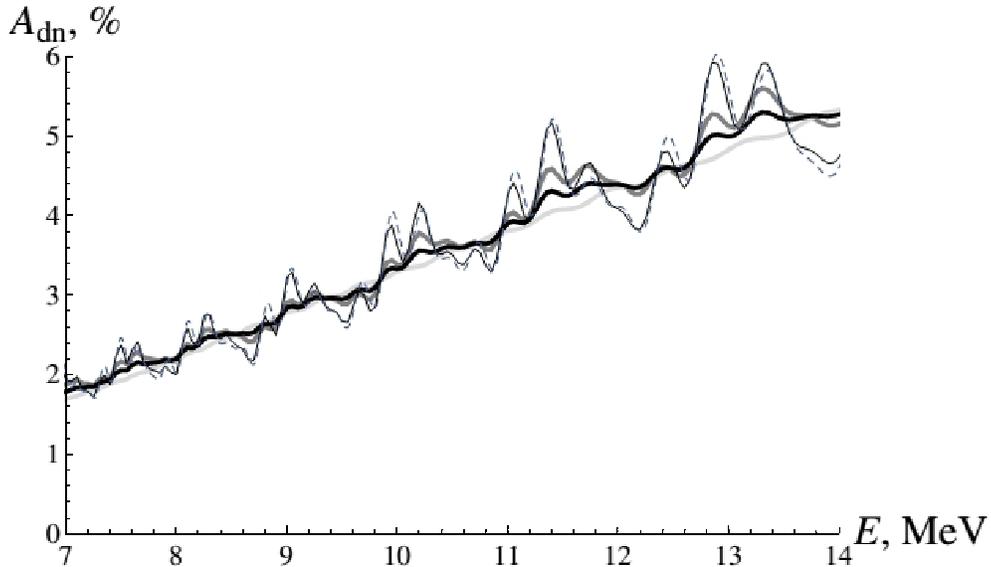}
\caption{Day-night asymmetry factor $A_{\mathrm{dn}}$ as a function of the neutrino energy $E$ for
the detector latitude $\chi = 26^{\circ}$ and different averaging seasons: Thick gray/lightgray
curves for the winter/summer half-years, thick black for the whole year, thin solid line
for December $+$ January. The dashed curve shows December $+$ January data for the Tropic
$(\chi = 23.5^{\circ})$ for comparison.}
\label{seas}
\end{figure}

In order to analyze the seasonal effects described above (which were in fact
discovered using analytical techniques) and to check the applicability domain
of the approximations used, we have made a numerical simulation. Namely,
we chose the weighting function $w(t) = w_0(e^{-\sigma^{2}/(t-t_{1})^{2}}+e^{-\sigma^{2}/(t-t_{2})^{2}}
\vartheta(t-t_{1})\vartheta(t_{2}-t)$ which smoothly limits the observation period to $t \in [t_1, t_2]$, and then
computed the time average \eqref{yearAvg_def}, using the numerical solution of the MSW equation.
FIG. 1 shows the day-night asymmetry factor $A_{\mathrm{dn}}(E)\approx \frac{T_{\mathrm{day}}}{T_{\mathrm{night}}}
\frac{\langle P_{\nu_{e}}({\mathrm{night}},E)\rangle}
{\langle P_{\nu_{e}}({\mathrm{day}},E)\rangle}- 1$, where $T_{{\mathrm{day,night}}}$ is the weighted total daytime/nighttime over a given season.

It is quite vivid from this picture that the amplitude of the `anomalous' oscillations
reaches as much as $20-30\% $ of the `trivial' effect and would be observable
at a detector with an improved neutrino energy resolution $E\sim 1-2$ MeV, say,
near S\~{a}o Paulo, Brazil $(\chi = 23.5^{\circ}S)$ \cite{5}. Thus, addressing the question raised in
the title above, we conclude that {\emph{one should look for spices near the Tropic, around the winter solstice}}.

\acknowledgments

The numerical simulations reported have been performed using the Supercomputing
cluster ``Lomonosov'' of the Moscow State University.


\begin{thebibliography}{9}

\bibitem{1} S. S. Aleshin, O. G. Kharlanov, and A. E. Lobanov, Phys. Rev. D {\bf 87},
045025 (2013).
\bibitem{2} P. C. de Holanda, Wei Liao, and A. Yu. Smirnov, Nucl. Phys. B {\bf 702}, 307
(2004).
\bibitem{3} S. Mikheev and A. Smirnov, Sov. J. Nucl. Phys. {\bf 42}, 913 (1985).
\bibitem{4} M. V. Fedoruk, {\it The Method of Steepest Descent} (Nauka, Moscow, 1977)
[in Russian].
\bibitem{5} In this case, the observations should be made around June, 22, since S\~{a}o Paulo lies in the southern hemisphere.
\end{thebibliography}
\end{document}